\documentstyle[12pt]{article}

\newcommand{\beq}{\begin{equation}}
\newcommand{\eeq}{\end{equation}}

\renewcommand{\thefootnote}{\fnsymbol{footnote}}
\def\aprle{\buildrel < \over {_{\sim}}}

\begin{document}
\topmargin 0pt
\oddsidemargin=-0.4truecm
\evensidemargin=-0.4truecm
\newpage
\setcounter{page}{0}
\rightline{IC/93/388}



\vskip 0.5truecm
\begin{center}
{\normalsize International Atomic Energy Agency\\
and\\
United Nations Educational, Scientific and Cultural Organization\\}
\vskip 1mm
{
INTERNATIONAL CENTRE FOR THEORETICAL PHYSICS}

\vskip 15mm

{\bf RESULTS FROM NEUTRINO EXPERIMENTS}
\footnote{Talk given at
the XVI International
Symposium ``Lepton-Photon Interactions",
Cornell University,\\
\phantom{iuyt} Ithaca, New York,  10 - 15 August 1993.}\\
\vskip 13mm
{Alexei Yu. Smirnov}
\footnote{On leave from Institute for Nuclear Research, Russian Academy of
Sciences, 117312 Moscow, Russia.\\
\phantom{okji}e-mail: smirnov@ictp.trieste.it}\\
{\it International Centre for Theoretical Physics, I-34100 Trieste, Italy}\\

\vskip 15mm
ABSTRACT
\end{center}
Recent (first or/and the best) results from the  neutrino experiments are
reviewed  and their implications for the  theory are discussed.
The sense of the experiments is the searching
for neutrino masses, mixing and interactions
beyond the standard model. Present laboratory experiments
give upper bounds on the masses and the mixing which are at
the level of  predictions of the ``electroweak see-saw".
Positive indications of nonzero lepton mixing follow from
studies of the solar
and  atmospheric neutrinos.

\vskip 2truecm
\begin{center}
MIRAMARE -- TRIESTE\\

\vskip 1mm

November 1993
\end{center}
\vspace{1cm}
\vspace{.5cm}
\renewcommand{\thefootnote}{\arabic{footnote}}
\setcounter{footnote}{0}

\newpage

\vskip 1cm

\begin{center}
INTRODUCTION
\end{center}

A ``Zoo" of the neutrino experiments includes:
direct measurements of the neutrino masses,
double beta decay searches,
oscillation experiments,
searches for the neutrino decays, ``peaks and kinks"
(in energy spectra of charged leptons from the weak decays),
detection of the atmospheric neutrinos,
spectroscopy of the neutrinos from Sun.
The main goal (or one of the main goals) of these experiments
is searching for the
neutrino masses and
mixing. Moreover,
the results of the experiments can be presented as certain
regions  (excluded, unexcluded, favoured, disfavoured, etc.) on the
unique plot of the mass and mixing parameters (say, $\Delta m^2$ and
$\sin^2 2\theta$ in two neutrino case).
Another goal is searching for  neutrino interactions beyond
the standard
model   (e.g., neutrino-Majoron coupling, electromagnetic interactions due
to large magnetic moment of neutrino, etc.).

An important information on the neutrino properties follows  from
``non-neutrino" experiments, (e.g., LEP measurements of the width of the
$Z^0$-boson), as well as from the astrophysics and cosmology
(evolution of stars, neutrino burst from SN1987A, primordial
nucleosynthesis, large scale structure of the Universe).

In a number of experiments a
neutrino is used as the ``probe" particle for  studies
of  structure and  interactions of other particles
(measuring the structure functions of the nucleons,
determination of the electroweak mixing angle and
the weak neutral
currents of the electrons, etc.)

In present review we
will consider the experiments which give the information on the
neutrino properties, and first of all, on the neutrino masses and mixing.\\

\begin{center}
FLAVORS AND MASSES
\end{center}

1. There are three neutrinos. Measurements of the invisible width of the
$Z^0$ boson
at LEP give the number of  light (m $\aprle$ 50
GeV) neutrino species with usual weak interactions :
\beq
N_{\nu} = 2.98 \pm 0.03,~~~ 90\% C.L.,
\eeq
(combined result of four collaborations$^1$).
With statistics increase, $N_{\nu}$ continuously converges to 3. The
bounds on
(N$_{\nu}$ - 3) have a number of applications to the neutrino physics. The
upper bound gives, e.g., the restriction on
the admixture of  the $SU_2$-nonsinglet state in the
Majoron field
and
consequently on the neutrino - Majoron coupling.
The lower bound restricts
the admixture of heavy ($m > m_Z$) neutral lepton, etc..

Three neutrinos correspond to three charged leptons and
the correspondence (i.e., flavor of neutrino) is established by
the charged current weak interactions.
Mass and flavor states may not
coincide. In the direct searches for  the neutrino
masses the data are fitted under  assumption that flavor state
has a definite mass
or that in flavor state only one mass
component dominates.
Let us describe  the data.\\

2. {\it Electron (anti)neutrino}: the bounds on  mass obtained from
tritium experiments$^{2 - 8}$ are shown in fig.1.
The best limit was announced by Mainz group$^{3}$. The result of
measurements,
$m^2 = -39 \pm 34 \pm 15$ eV$^2$  ($1 \sigma$), gives
\beq
m(\bar{\nu}_e) <   7.2~ {\rm eV},~~ 95\%~ C.L.~ .
\eeq
The value of $m^2$ is about $1 \sigma$ below zero
which  may testify for some undiscovered systematic error,
especially in view of the fact that  all other experiments also give
negative values of $m^2$. (For discussion of this problem see$^{8}$).
In this connection,
the discrepancy was pointed out$^{3,6}$ between the data and the spectrum
expected
from the theoretical final state distribution. In 1993,  Mainz group has taken
new data for 3 months  and it is expected that statistics accumulated
can considerably improve the limit (2),  as well as can help to understand
possible
systematic errors. Also for this purpose the measurements are performed with
conversion electrons from $^{83m}$Kr.\\

3. {\it Muon neutrino}: the best limit on the mass follows from
 study of  the pion decay  $\pi^+
\rightarrow \mu^+ \nu_{\mu}$  at the rest. Two values are involved:
the momentum of the muon which has been measured recently with better
precision  at PSI$^{9,10}$
and from independent measurements of  the pion mass.
Method of  $m_{\pi}$
determination
has an ambiguity:  two different values of $m_{\pi}$ where obtained.
One  value  results in  strongly negative $m^2(\nu_{\mu})$
(about $5\sigma$ below zero),  another one  gives
$m^2 (\nu_{\mu}) = - 0.002 \pm 0.030~ {\rm MeV}^2$ in agreement with zero.
The latter corresponds to  a new upper bound$^{10}$
\beq
m(\nu_{\mu}) < 220 ~{\rm kev}, ~~~ 90 \% ~C.L.~.
\eeq

4. {\it Tau neutrino}. The invariant masses of five pions in
the  decay
$\tau \rightarrow 5 \pi  \nu_{\tau}$
are measured
and an upper bound on the tau neutrino mass follows from spectrum of the
invariant masses
near the  end point:
\beq
m(\nu_{\tau}) <  \left\{ \matrix{ 31~ {\rm MeV} ~~ 95 \%~ C.L. & {\rm
ARGUS}^{11} \cr
                             32.6~ {\rm MeV} ~~ 95 \%~ C.L. & {\rm CLEO}^{12}
\cr  } \right. .
\eeq
The electromagnetic calorimetry allows CLEO
to detect the decays with neutral pions: $\tau \rightarrow (3h)^-  2\pi^0
\nu_{\tau}$. As much as 53 events of such a type have been found.
A sample analyzed includes also  60 events with 5
charged pions.
The limit is determined, actually,
by a few  events near the end point.
ARGUS group detecting only charged pions is lucky: the limit was obtained
with just 20 events; the probability to get such a limit  is$^{12}$ $p \approx
0.04\%$. The probability of CLEO limit is 13.9 \% .\\

5. Strong bounds on tau neutrino mass follow from
a {\it primordial nucleosynthesis}$^{13 - 16}$ (fig.2).
At the epoch of nucleosynthesis the neutrinos with
masses $m > 0.1$ MeV were nonrelativistic
(in appreciable  part of spectrum).
Their contribution to the energy density
in the Universe is characterized by  $N_{eff}$ -- the effective
number of the relativistic neutrinos giving the equivalent
amount of
energy.  The contribution of massive neutrinos is
the interplay of two factors: $\Delta N_{eff} \propto m \cdot n(T)$,--
the  mass,  and the concentration, $n(T)$;
the latter is exponentially suppressed at temperatures $T < m$. As the result
of the interplay the energy density,
has a maximum at $m$ = (4
- 6) MeV, and in maximum:  $\Delta N_{eff} \cong (4 - 5)$.
Total effective number of the relativistic neutrinos,
$N_{eff}$, is
restricted by present $^4$He abundance.
The limit $N_{eff} < 3.4$ was used in  original paper$^{13}$  and the
regions of  masses of the Majorana neutrino
(0.5 - 25) MeV  and (0.5  - 32) MeV were excluded for the
neutrino lifetimes larger than 1 s and  $10^3~$s correspondently.
Recently the bounds have been refined.
In$^{14}$  using even  weaker restriction,
$N_{eff} < 3.6$,
the region 0.5 - 35 MeV was excluded for $\tau > 10^2$ s.
The limit $N_{eff} < 3.3$ forbids$^{15, 16}$ the interval
0.1 - 40 MeV for
$\tau_{\nu} > 10^3~ $s.
Consequently, for stable neutrinos  there is no gap
between the laboratory (4) and NS bounds and the upper limit is pushed down
to  \beq
m(\nu_{\tau}) < 0.1~ {\rm MeV},~~~   (\tau > 10^3 s ).
\eeq
Fast decay
$\nu_{\tau} \rightarrow \nu ' + \chi$, where $\chi$ is the Majoron, can
relax the bound. It has been found$^{16}$ that the gap appears for
$\tau_{\nu} < 10^{2}~ $s.
For m = 30 MeV the lifetime $10^{2}$s corresponds to
the nondiagonal neutrino-Majoron coupling constant $g_{\chi}' \sim
10^{-12}$.
 At $\tau_{\nu} < 10^{-2}~$s the mass region  3 - 30 MeV
is not  excluded.
The gap for stable neutrinos appears if one admits $N_{eff} > 4$.\\

6. The above result has a number of implications.
A relatively  stable tau neutrino should be appreciably lighter than
the electron;  there is no decay
$\nu_{\tau} \rightarrow e^+  e^- \nu '$. If the
mass of $\nu_{\tau}$ is generated by the see-saw mechanism, the
corresponding Majorana mass of the right handed component should be of the
order of $m_{\tau}^2/ m(\nu_{\tau}) \approx 3 \cdot 10^4$ GeV,
which is appreciably larger than the electroweak scale, i.e., the
electroweak see-saw does not work, etc..

It is important to strengthen the laboratory upper bounds:
the discovery of the tau neutrino mass in the region about 30 MeV
will mean either that the neutrino is unstable, and  moreover, the
invisible modes (like the  Majoron one) should dominate,  or that
present picture of the primordial nucleosynthesis is incorrect.

The above nucleosynthesis limit is applied also to the muon
neutrino.\\

7. Let us define the following mass scales:
\beq
m_1 = \frac{m_e^2}{M} \equiv 3.2~ {\rm eV} \frac{m_W}{M},~~~
m_2 = \frac{m_{\mu}^2}{M} \equiv 135~ {\rm kev} \frac{m_W}{M},~~~
m_3 = \frac{m_{\tau}^2}{M} \equiv 39~ {\rm MeV} \frac{m_W}{M},
\eeq
where $m_e$, $m_{\mu}$, $m_{\tau}$ and $m_W$ are
the
masses of the electron, muon, tau lepton and W-boson correspondently.
Such relations can
arise from the see-saw mechanism of mass generation with  Majorana mass of
the  right handed neutrinos m($\nu_R$) = M. At
M = $m_W$ the masses in (6) coincide (up to the factor of 2)   with
present upper bounds on the neutrino masses (2 - 4). This means that  the
sensitivity of present searches is at the level of the electroweak
see-saw (M = 30 - 300 GeV).
(Although it is unclear how the singlet $\nu_R$ ``feels" this
scale).

Physics of solar neutrinos determines another scales:
\beq
m_{\odot} = (10 ^{-5} - 3\cdot 10^{-3}) {\rm eV},
\eeq
which imply much higher values  of M.\\

\newpage

\begin{center}
DOUBLE BETA DECAY SEARCHES
\end{center}

1. Two neutrino double beta decay, $(\beta\beta)_{2\nu}$:
$ Z \rightarrow (Z + 2) + e^- + e^-  + \bar{\nu} + \bar{\nu}$
was observed already for five
nuclei:
$^{76}$Ge,
$^{100}$Mo,
$^{130}$Te,
$^{82}$Se,
$^{238}$U.
Its study gives to some extend the ``calibration"  of  nuclear
matrix elements.
Searches for  the neutrinoless  mode,
$(\beta \beta_{0\nu})$,
$
Z \rightarrow (Z + 2) + e^- + e^-
$
are sensitive, in particular,  to the effective
Majorana mass of the  electron neutrino:
\beq
m_{ee} \equiv \sum_{i} \eta^{CP}_{i} |U_{ei}|^2 m_i,
\eeq
where
$U_{ei}$, $m_i$, and $\eta^{CP}_i$ are the admixture
in the electron neutrino state, the mass and the CP- parity
of the i-component of neutrino ($m_i < 30$ MeV).
(In case of opposite parities the cancellation
in (8) allows even for lightest component to have a mass
above the upper bound on $m_{ee}$).
The Majoron mode of the decay, $(\beta\beta)_{0\nu \chi}$,
$
Z \rightarrow (Z + 2) + e^- + e^- + \chi,
$
being a signature of the spontaneous violation of the lepton number
symmetry, determines
the neutrino Majoron - coupling, $g_{\chi} = m(\nu)/\sigma_0$,
where $m$ is the neutrino mass and
$\sigma_0$ is the scale  of lepton
number violation.\\

Let us describe the results of experiments.

2. {\it Heidelberg - Moscow} Collaboration (Gran Sasso
underground laboratory)$^{17 - 19}$ performed the experiments with
three $^{76}$Ge - detectors. Total active mass of the enriched (86
\%) isotope
is  6.01 kg; total exposure time is 6.2 kg$\cdot$y (70.75 mol*y). (Fourth
detector with mass 2.88 kg is available).

In a study of the
$(\beta \beta)_{2\nu}$
-mode the Collaboration used the data
taken by the detector \# 2 (2.6 kg)  from September 1991 to August
1992.
In the interval 800 - 1500 kev
after background subtraction (the model of the background build from
measured quantities)
more than 4100 events have been
prescribed to $(\beta \beta)_{2\nu}$ decay.
The  energy distribution of two electrons
as well as the half life,
$
T_{1/2} = (1.42 \pm 0.03 \pm 0.13) \cdot 10^{21} y, ~ (90 \% C.L.)
$
agree with expectations, and the error bars reflect the amusing progress
in the field.

In searches for the neutrinoless mode
all the  exposure time was used and the best
limit$^{17}$  has been obtained
\beq
T_{1/2} > 1.5~ (2.4) \cdot 10^{24}~ y, ~~~ 90 (68)\% C.L.~.
\eeq
It corresponds to the effective Majorana mass
\beq
m_{ee} < 1.3~eV, ~~~90\% C.L.
\eeq
for ``Heidelberg" nuclear matrix element.
These results, however,  were not changed practically during
last year although the exposure time  has increased appreciably.
The reason is the appearance  of the excess of
events at the end point. The following features of the excess are
remarked: (1) it has a shape of the peak at the energy E =
2040  kev which coincides with  the end point; (2) the width
of the
peak $\Delta E = 4.4$ kev corresponds to the energy resolution of the
detector, 3.7 kev; (3) the
increase of
the number of events in the peak with time can be described by linear
dependence; (4) the number of the events in the peak ($\Delta E = 8.8$
kev) equals $N_p = 28$, whereas the expected number of the
background events (extrapolation of the signal from the regions outside
the peak) is $N_{b} = 15$. The peak can be explained as $2.7 \sigma$
fluctuation of the background; the fit of the data by using the
hypothesis ``background plus  peak" gives
$1.5 \sigma$ significance of the peak.
In August'93 new sample of the data has
been analyzed$^{19}$, exposure time has reached 83.03 mol*y.
Since the background increased more fastly than the peak did, the
significance of the excess has decreased from 2.7 to 2.2$\sigma$.
The interpretation
of the effect is still unclear, but
the excess is strong enough to be proved or disproved within
reasonable time scale ($\sim$ 1 year).

The Majoron mode of the decay was searched$^{18}$ in
energy ``window" 1100 - 2050 kev (71\% of all expected
events). Total number events, 208,  has
been found, whereas  the background model
gives $N_{\sigma} = 92.5$ events. Thus the $2.25\sigma$
excess
has been observed , but the energy distribution of events has a
``wrong shape"$^{17}$. The limit
$T_{1/2}(M^0) > 1.7 (1.9) \cdot 10^{22}$ yr
(90 (68) \% C.L.) has been obtained which gives
the restriction on the Majoron coupling:
$g_{\chi} < 1.8 \cdot 10^{-4}$, 90\% C.L. .\\

3. {\it Neuchatel-Pasadena-PSI} Collaboration (Gottard underground
laboratory) has published$^{20}$ the results of study of the $^{136}$Xe decay
with xenon time projection chamber. Active volume is 180 liters of the
enriched (62.5\%) Xe at 5 atm.. Data taking for 6830 h gives the bound on
half-life for neutrinoless (mass) mode:
$T_{1/2}(0\nu) > 3.4~(6.4) \cdot 10^{23}$ years, 90\% (68\%) C.L. .
This corresponds to the upper bound $m_{ee} < 2.8 - 4.3$ eV.
The estimations of the
nuclear matrix
elements for $^{136}$Xe have no large spread, in contrast with case of
$^{76}$Ge. For the Majoron mode the
limit $T_{1/2}(0\nu \chi) > 4.9 \cdot 10^{21}$ years has been obtained;
it gives for the Majoron coupling: $g_{\chi} < 2.4 \cdot 10^{-4}$.
Collaboration plans to reduce the background and to reach the sensitivity
$\sim 10^{24}$ yr for $(\beta \beta_{0\nu})$ mode.\\

4. {\it NEMO} Collaboration (Frejus underground laboratory) has published
first results of searching for the double beta decay of the $^{100}$Mo
with the detector NEMO-II$^{21}$. The detector consists of tracking volume
(frames of Geiger cells);  the source foil placed in central frame has two
parts: one contains 172 g of enriched (98.3 \%) isotope $^{100}$Mo, and
second one -- 163 g of natural (9.6\%) isotope. On the opposite sides of
the tracking volume there are two scintillator walls which allow to make
the energy and the time flight measurements.  The data where collected
from December 92 to May 93 with total exposure time 2485 h.  The
subtraction of the results (enriched - natural) has been done to remove
the external background. After the subtraction the sample consists of 454
events; their energy distribution is perfectly described by the expected
one from the $2\nu$ decay. The obtained lifetime:  $T_{1/2} > (1.0 \pm
0.08 \pm 0.2) \cdot 10^{19}$ years, agrees with previous results.  No
events have been observed above 2600 kev (the end point is 3030 kev) which
gives the upper bounds on the neutrinoless mode:  $T_{1/2}(0\nu) > 3.8
\cdot 10^{21}$ yr , as well as on the Majoron mode:  $T_{1/2}(0\nu \chi) >
5 \cdot 10^{20}$ yr (90 \% C.L.).  The latter result corresponds to rather
strong limit on the Majoron coupling:  $g_{\chi} < 1.8 \cdot 10^{-4}$. At
present (October'93)$^{22}$ the exposure time has increased up to 6140 h
(1.2 mol*y), the number of $(\beta\beta)_{2\nu}$ events is 1302
(background subtracted), which gives $T_{1/2}(2\nu) > 1.1 \pm 0.03(stat)
\cdot 10^{19}$ yr. New limit on the Majoron mode is $T_{1/2} > 1.7 \cdot
10^{21}$ yr, 90\% C.L.  which strengthens the bound on $g_{\chi}$ by factor
of $\sim 1.5$.

{\it LBL-MHC-UNM-INEL} Collaboration$^{23}$ has improved their limit on
the half life of the neutrinoless double beta decay of $^{100}Mo$ by
factor of 11. The Si(Li) detector is used; the mass of isotope is 60.63 g;
3849.5 h of exposure time give $T_{1/2}(0\nu) > 0.44 \cdot 10^{23}$ years
which corresponds to $m_{\nu} < 6.6$ eV 68\% C.L. .\\

5. {\it Washington-Tata} group has studied the double beta decay of
$^{128}$Te and $^{130}$Te (Te $\rightarrow$ Xe) by the geochemical
method$^{24}$. The
ancient
Te - ores ($10^9$ years) were used and the Xe-atoms produced in the decay
were detected by the ion-counting mass spectroscopy.
The double beta decay is considered
to dominate in the production of Xe in ores.
Thus the measured
ratio of $^{128}$Xe and  $^{130}$Xe concentrations
 gives the ratio of the half lifes:
\beq
\frac{T_{1/2}(^{130}Te)}{T_{1/2}(^{128}Te)} = (3.52 \pm 0.11) \cdot
10^{-4},   \eeq
and this ratio agrees well with
two neutrino decay mode. Using Pb-dating one  finds the absolute value
of the half life for $^{130}$Te:
$T_{1/2}(^{130}Te) = (2.7 \pm 0.1) \cdot 10^{21}$ yr,
and consequently, using the result (11):
$T_{1/2}(^{128}Te) = (7.7 \pm 0.4) \cdot 10^{24}$ yr.
The ratio (11) allows also to get the upper bounds on the neutrinoless
modes, and consequently, on the Majorana mass:
$m_{ee} < (1.1 - 1.5)$ eV. Suggesting that all the $^{128}$Te - decays
are due to the Majoron mode one gets the bound on the Majoron coupling
$g_{\chi} < 3 \cdot 10^{-5}$.\\

6. Let us comment on the implications of the results.
The bounds on $m_{ee}$ and $g_{\chi}$ from the discussed
experiments are summarized in (12)
\beq
\begin{tabular}{|l|l|l|l|}
\hline
\sl Experiment  & \sl Element & \sl $m_{ee}$ $<$ , {\rm eV}, 90 \% C.L. &
\sl $g_{\chi} \cdot 10^4$ $<$ \\
\hline
Heidelberg-Moscow & $^{76}Ge$  &  1.3            &  1.8 \\
Neuchatel-Pasadena-PSI  & $^{136}Xe$ &  2.4 - 4.3  &  2.4 \\
NEMO-II           & $^{100}Mo$ &  7                &  1.8 \\
Tata-Washington   & $^{130}Te$ &  1.1 - 1.5        &  0.3 \\
\hline
Nucleosynthesis   &    -     &   -                &  0.09 \\
\hline
\end{tabular}
\eeq
Present limits on the effective Majorana mass are at the level
(1 - 2) eV. In future Heidelberg-Moscow Collaboration will perform the
experiment with 20 kg of the enriched
$^{76}$Ge; a similar amount of the enriched $^{76}$Ge will be used in
IGEX experiment$^{25}$;  NEMO collaboration
intends to use 10 kg of the enriched $^{100}$Mo (also the experiments
with other enriched isotopes are planed). The sensitivity
of these experiments to the neutrino mass will reach  0.1 - 0.2 eV,
and the mass interval $m_{ee} = 0.1 - 1$ eV will be ``observable".\\

7. What are the implications of new searches? A straightforward
interpretation of the positive signal in $\beta\beta_{0\nu}$ is that the
electron
neutrino consists mainly of the  lightest mass eigenstate, $\nu_1$,
having the mass in the indicated region.  The mass $\sim$ 1 eV
can arise from the see-saw mechanism at the electroweak scale.
For masses of two
other
components one  predicts then $m_2 = (1 - 100)$ kev and $m_3$ = (1 - 30) MeV.
However, this scenario does not allow to
solve  the solar and  atmospheric neutrino problems by the neutrino
oscillations or resonant conversion. The
conversion $\nu_e \rightarrow \nu_{\mu}$ of solar neutrinos
implies  $m_2 \approx m_{\odot} \approx 3 \cdot 10^{-3}~$eV, then
according to the see-saw the mass
of third neutrino can be in the region (1 - 30) eV. But  its
admixture
to the $\nu_e$ state is typically predicted to be very small
$\left(|U_{e3}|^2 < \frac{m_1}{m_3} <
\frac{m_{\odot}}{m_3} \right)$, and consequently, the contribution to the
effective Majorana mass is negligible: $m_{ee} \sim m_{\odot}$.
There are several ways to reconcile the ``observable" Majorana mass
and the neutrino physics solution of the $\nu_{\odot}$-problem$^{26}$.
The admixture of the $\nu_3$ in the $\nu_e$ state can be enhanced
so that the effective Majorana mass is due to admixture of the third
neutrino: $m_{ee} = |U_{e3}|^2 m_3$ (fig. 3).
The enhancement  can be obtained by the see-saw mechanism with
certain structure  of  mass matrix of
the RH neutrinos. In turn,
this structure can be a consequence of certain family symmetry at high
mass scales and it implies a strong mass hierarchy of the RH
neutrinos.

Another extreme case corresponds to
strongly degenerate spectrum:
$m_1 \approx m_2 \approx m_3 \sim m_{ee}$. Small mass splitting
$(m_2 - m_1) / m_{ee} \sim m_{\odot}^2 / m_{ee}^2$
allows to solve
the solar neutrino problems, and for larger splitting
$m_3 - m_2$ -- to solve the atmospheric neutrino problem.
The degeneracy can follow from horizontal symmetry,
whereas small splitting and mixing result$^{26}$ from radiative corrections or
from the see-saw contribution or from  Planck scale effects
(see below and$^{90,91}$).\\

8. Direct searches give the upper bound on the Majoron coupling at the
level $(1 - 2) \cdot 10^{-4}$; more strong bound follows from the
geochemical experiment:
$g_{\chi} < 0.3 \cdot 10^{-4}$. Primordial
nucleosynthesis gives even more strong restriction$^{27}$. The
contribution
of Majorons to the  energy density in the Universe has been calculated and
the upper bound on number of the relativistic degrees of  freedom,
$N_{eff} < 3.3$, allows one to get
$g_{\chi} < 0.09 \cdot 10^{-4}$.
These results strongly disfavour the
interpretation of the excess of  events observed in the experiments with
three different nuclei$^{28}$
in terms of  usual Majoron decay.\\

9. It is possible to construct the models$^{29}$
 with large neutrino-Majoron
coupling ($10^{-5} - 10^{-4}$)
which do not contradict to  LEP bound (1).
But in this case the upper bound on $m_{ee}$
implies that the  scale of lepton number violation
is as small as $\sigma_0 = 10 - 100$ kev.
Such a scale can be naturally protected by some kind of supersymmetry$^{30}$.

The double beta decay with emission of massless scalar particle may take place
without  lepton number violation, so that scalar carries  double lepton
charge and the Majorana
neutrino mass is zero$^{31}$. Such a scalar may appear as the Goldstone boson
at  spontaneous violation of some new symmetry which is not related to the
lepton number. The scale of violation can be as large as 100 MeV which
allows one to escape from the Nucleosynthesis bound.
The double beta decay with ``charged Majoron"
has more soft   energy spectrum
of two electrons than the standard Majoron decay$^{31}$.

It was argued that all global symmetries are broken by gravity (Planck
scale interactions) which means that massless Majoron does not exist
at all$^{32,33}$.
At low energies the effects may be described by nonrenormalizable effective
interactions with can drastically change the picture of the lepton number
violation for small scales $\sigma_0$. For example, the term
$\lambda \frac{\Phi ^4 \sigma}{M_{Pl}} + ...$, where
$\Phi$ is the usual Higgs doublet with vacuum expectation $v$,
could generate the mass
$m_{\chi}^2 \sim \frac{v^4}{M_{Pl} \sigma_0} \sim 1$ MeV of the order of the
energy release in the $\beta \beta$ decays.\\

\begin{center}
``KINKS, PEAKS, DECAYS". OSCILLATIONS
\end{center}

1. Vacuum mixing implies that  flavor neutrino states are
composed of
several states with definite masses, e.g., the electron neutrino
is $\nu_e =  \sum_{i} U_{ei} \nu_i.$
Mixing has a number of consequences:
kinks on the Kurie plot of beta decays,
additional peaks in energy distributions of charged leptons from two
body decays, for example,  $\pi \rightarrow \mu \nu_{\mu}$,
neutrino decays, oscillation of neutrinos, etc..\\

2. {\it Kinks}. No kinks on the Kurie plots have been found
in recent high statistics and high precision experiments$^{34 - 38}$
in kev region.
This gives the  upper bound on mixing parameter $|U_{ei}|^2$
as function of neutrino mass
\beq
 \begin{tabular}{|l|l|l|l|l|}
   \hline
\sl Experiment & \sl Isotope & \sl $U_{eh}^2$ $<$ & \sl C.L.& \sl mass, kev
\\
\hline
INS ~{\rm Tokyo}$^{34}$ & $^{63}Ni$  &  0.073 \% & 95\% & 17 \\
 --''--               & $^{63}Ni$  &  0.15 \%  & 95\% & 10.5 - 25~\\
{\rm Z\"urich}$^{35}$   & $^{63}Ni$  &  0.11 \%  & 95\% & 17 \\
{\rm Argonne}$^{36}$  & $^{35}S$   &  0.25 \%  & 95\% & 10 - 45\\
{\rm Princeton}$^{37}$ & $^{35}S$   &  0.29 \%  & 95\% & 17\\
{\rm Oklahoma}$^{38}$   & $^{3}H$    &  0.24 \%  & 99\% & 17\\
\hline
\end{tabular}
\eeq
As it was noted by A. Hime$^{39}$,  these recent experiments
``definitely ruled the presence of a 17 kev neutrino and circumvent the
criticisms applicable to earlier ``null" results".
More subtle question is what is the origin of spectra distortion in the
``non null" experiments? Recent  studies (see for review$^{39}$) show  among
the reasons, e.g., the electron scattering effect on the way
from the source to detector.
Nevertheless the
kev region is  interesting. The electroweak see-saw
gives the mass of the second neutrino in this region. Mixing
parameter could be as small as 0.1 - 1 \%.
Models
developed in context of the 17 kev neutrino
(in particular with radiative generation of masses)
predict more naturally  smaller
mixing than it was found in the
``positive"  experiments (0.8 - 1.2 \%). One may keep in mind $|U_{eh}|^2 \sim
m_e/m_{\tau}
\approx 0.03 \%$ or $(m_{\mu}/m_{\tau})^2 \approx 0.25 \%$, etc.. \\

2. {\it Neutrino decays}. If the electron neutrino has an admixture
$U_{eh}$
of state $\nu_h$ with mass $m_h > 1~$ MeV, the
decay $\nu_h \rightarrow e^+ e^- \nu_e$  takes place. The bound on
the  lifetime of $\nu_h$, and consequently, on $U_{eh}$
has been improved recently by {\it Munich-Annecy-Marseille} group$^{40}$.
The decay was searched for in the antineutrino beam from the reactor
BUGEY. The detector placed at  distance 18.6 m from the core
of reactor consists of the He-filled decay volume $ \sim 2 \times 2 \times 2$
m$^3$ and the electrons are detected by position sensitive multiwire
proportional chambers, placed at the opposite (to  reactor)
side of the detector. No decays have been observed during the run of the
experiment in 1991. The upper bound on the decay rate ( $ < 0.012~$
 s$^{-1}$,  90 \% C.L.)  gives the best limit
$|U_{eh}|^2 < 2 \cdot 10^{-4}$ (90 \% C.L.)
in the region (3 - 6) MeV.
This results improve the previous limits from ``G\"osgen-87" and ``Rovno-90"
by factor of 3. The analysis of results from next run of the experiment
will allow to improve the  limit up to    $(5 - 7) \cdot 10^{-5}$.

Some remarks are in order.
The $\nu_h$ could be
a main component of the tau neutrino. In this case the region of
sensitivity  of the BUGEY experiment, $m_h  \sim 1 - 10$ MeV,
 is strongly disfavoured by Nucleosynthesis, unless the neutrino has
some other decay mode like a Majoron one with $\tau < 1$ s.
The
$\nu_{\tau}$
as the Dirac
neutrino is excluded by  data from SN87A$^{41}$ (the upper bound is
about 20 - 30 kev). If $\nu_{\tau}$ is the Majorana particle, then
strong bound on mixing follows from the
$\beta\beta_{0\nu}$ searches:
$|U_{e\tau}|^2 < m_{ee}/ m_h < 2 \cdot 10^{-6} (m_h/MeV)^{-1}$,
where for $m_{ee}$ the upper bound (10) is used.
The bound is much stronger than the existing and  planning limits.
To avoid it one should suggest strong cancellation of the
contributions in $m_{ee}$.
The possibility of $\nu_h$ to be a sterile neutrino is disfavoured by the
primordial nucleosynthesis consideration. \\

3. {\it L-3 collaboration}  (LEP) searches for isosinglet neutral heavy
lepton, $N$, that mixes with active neutrino states$^{42}$. If $m_N < 90$
GeV   then $Z^0$ decays into $\nu N$, and $N$, in turn, decays into
$lq \bar{q}$,
$\nu q \bar{q}$,
$\nu l \bar{l}$ . A signature, e.g., of the first mode is missing energy,
charged
lepton, and two jets. The data are compared with Monte Carlo predictions.
No excess of the events above the background has been found which gives
the upper bound on
$|U_{lh}|^2$ as  function of the lepton mass.
Preliminary result
for
the $N$ admixture in, e.g.,  the $\nu_{\tau}$
is: $
|U_{\tau h}|^2 <  (0.7 - 1.0) \cdot 10^{-4}$ ($m_h = 5 - 60~ $GeV)
(90 \% C.L.).

The result is important for the electroweak see-saw. If the mass of the
right handed neutrino is $\sim$ 60 GeV, then its admixture in the tau
neutrino state may be as large as $(m_{\tau}/m_N)^2 \sim 10^{-3}$.\\

4. {\it Oscillations}.
The reactor experiment {\it BUGEY-III}$^{43}$ is essentially accomplished,
and the data are  analyzed.
Preliminary
results of measurement of the $\bar{\nu}_e$ spectra at the distance
95 m from the reactor are published. During the run of
1992, about 1200 neutrino events have been detected and no
effects of oscillations have been found. In particular,
the ratio of signals
measured by the same detector from two different reactors (95 m/ 15 m)
does not depend on the neutrino energy.
New regions of the neutrino parameters can be excluded
in comparison with the existing results (fig. 3). Thus
G\"osgen bound on $\sin^2 2\theta$ will be improved up to the factor of
1.5 - 3 in the region $\Delta m^2 = (3 \cdot 10^{-2} - 5 \cdot 10^2)$
eV$^2$. These
new limits follow mainly from a comparison of spectra from two distances
15m/40m (the same detector) and from measuring of signal at 15 m.\\

5. {\it CHARM-II} collaboration has published the  limit on $\nu_{\mu} -
\nu_{\tau}$ oscillations$^{44}$.
In the detector $\nu_{\tau}$  would produce the $\tau$-leptons
that decay, in particular, as   $\tau \rightarrow \nu_{\tau} \pi$.
No excess of the
events with single pion
has been observed. The limit on $\sin^2 2\theta$ is only
factor of two weaker
than the best limit on this mode
from E531.

There are new results from E645 oscillation experiment$^{45}$ at meson
factory LAMPF.
For ``non-exotic" mode $\bar{\nu}_{\mu} \leftrightarrow \bar{\nu}_e$
the upper bounds have been  found
$\Delta m^2 < 0.14$ eV$^2$
at maximal
mixing and $\sin^2 2\theta < 0.024$ for large  $\Delta m^2$.

Essential progress in the field will be related to new high precision
experiments
CHORUS$^{46}$ and NOMAD$^{47}$ at CERN which will start next year,
as well as to long base line experiments. \\

6. It has been argued in$^{48}$ that
for the $\nu_{\tau}$
mass in the cosmologically interesting domain 2  - 30 eV
the strong restrictions on  the $\nu_{\tau} - \nu_e$ mixing can be obtained
from Supernova. Such a mixing will induce the resonant conversion of the
$\nu_{\tau}$ in the
$\nu_e$
near the core of the star thus producing
$\nu_e$ with high energies (original $\nu_{\tau}$ have about two times
higher average energies than  $\nu_e$). High energy electron neutrinos
will strongly suppress r-processes. If supernovae are produce
r-process heavy elements, then region
 $\sin^2 2\theta > 10^{-4} - 10^{-5} $ is excluded.\\

\begin{center}
ATMOSPHERIC NEUTRINOS
\end{center}

The atmospheric neutrinos are formed in the decays of  pions:
$\pi \rightarrow \mu \nu_{\mu} \rightarrow e \nu_{\mu} \nu_e \nu_{\mu}$
(and, in a smaller part, of kaons).
Pions, in turn, are
generated in
interactions of cosmic rays with nuclei of atmosphere.

Several types of events induced by the atmospheric neutrinos are studied
in the  underground detectors (for details see$^{49}$).
{\it Contained events}: neutrinos interact in a fiducial volume of
detector producing in quasielastic scattering  the electrons and muons;
(also in some part of events pions are produced).
The electrons and muons show up as  the ``e-like" (diffuse rings, showers)
and ``$\mu$-like" events (sharp rings, tracks) correspondently.
The trajectories of the secondary particles
are contained completely or partly in the detector.
The energies of these events are 0.2 - 1.5 GeV.

{\it Upward going muons}: Muon neutrinos produce the muons in the rock
that surrounds
the detector. The time and the angular resolution of detectors allow to
pick up the muons arriving from the down semisphere (also horizontal
muons were studied). These events, in turn, are divided into two
categories: {\it stopping upward going} muons  (muons  decay in the
detectors) and {\it through going} muons.

Typical energies of the original neutrinos are:
5 - 10 GeV,
20 - 100 GeV, and 50 - 300 GeV
for contained events, stopping muons, and through going muons
correspondingly.\\

2. The atmospheric neutrino problem is formulated  as the deviation
(smallness) of  the  {\it double ratio}
\beq
R(\mu/e) = \frac{(\mu-like)/(e-like)_{data}}{(\mu-like)/(e-like)_{M-C}}
\eeq
for the {\it contained} events  measured by
water \`{C}erenkov detectors Kamiokande and IMB  from 1 $^{49 - 52}$:
R($\mu$/e)
is about  0.6. The results are summarized in fig. 4.

The latests {\it Kamiokande} results further confirm the smallness
of R($\mu$/e). The data  correspond to the observation up to July
'93: total exposure time is 6.18 kt*y. For visible energies
$E^{vis} < 1.33$ GeV,
557 fully contained single ring events have been observed$^{50,51}$. Among
them there
are 191 $\mu$-like and 198 e-like events. The ratio $\mu/e = 0.96$
should be compared with  1.60 - 1.63,
predicted by different groups.
As the result for the double ratio one finds
\beq
R(\mu/e) = (0.59 - 0.60) \pm 0.06(stat) \pm 0.05 (syst), ~~(1 \sigma).
\eeq
The absolute number of the observed events is (0.74 - 1.12) of the
predicted value. The shape of energy distribution of the
events  is in
agreement with expectation. (Although one can remark the
excess of the events
in the the energy bit 0.4 - 0.5 GeV and the deficit in the bin 0.5 - 0.6
GeV both in e- and in $\mu$- spectra).

It is of great importance for implications  to study the effect at
higher energies.
For $E^{vis} > 1.33$ GeV Kamiokande has 110 fully contained events
(7.33 kt*y)and 89 vertex contained events. Among them there are 116
$\mu$-like and 83 e-like events and it seems the ratio $\mu/e $
is smaller than  expectation too$^{51}$.

Similar results have been obtained for the contained events
by IMB Collaboration$^{52}$.

There is a consensus that the uncertainties in the predictions of the
ratio $(\mu- like)/(e-like)$  are smaller than 5\%.
(The spread of values predicted  by different authors is even
smaller).
The misidentification of
the events in
water \`{C}erenkov detectors has been discussed as one of possible reasons
of small double ratio. But well identified events with muon decays
(signal from  decay electrons is detected) confirm the deficit of muon
neutrinos. Moreover,  the
calibration experiment is planning at KEK$^{53}$ to check possible
methodical effects. \\

3. Iron calorimeters Frejus$^{54}$ and  NUSEX$^{55}$ do not show the
anomaly (see fig.4). Measured value of the double ratio
agrees  with 1, although the errors are rather large.
Does this testify for methodical origin of the  the atmospheric neutrino
problem?
In this connection new results of SOUDAN-II Collaboration are of great
interest.

SOUDAN-II detector is the
iron calorimeter with tracking drift chambers.
The experiment had been started in April 1989 with 275 tons and then the
mass of the detector was increased, being 680 t in August 91, and
900 t since July 93 ; a complete mass, 950 tons, is planning to be
in fall of 1993. First publication corresponds to 0.5 kt*y recorded
up to August '91.
Now an additional 0.5 kt*y has been analyzed$^{56,57}$.
The total sample ($\sim$ 1 kt*y) consists of 579 fully
contained events. After  energy cut ($E > 200$ MeV) one finds  98
candidates
 72\% of which are quasielastic events (one charged lepton).
These raw data contain 34 tracks ($\mu$-like) and 32 showers (e-like
events).
The correction due to shield inefficiency gives  33.5 track and 33.3
showers. This should be compared with predictions (Monte Carlo): 42.6
tracks and
29.1 showers. The double ratio (``provisional"
result)  is
\beq
R(\mu/e) = 0.69 \pm 0.19(stat) \pm 0.09 (syst),~~~ 1 \sigma
\eeq
which is  larger than the result from the first 0.5 kt*y:
R($\mu/e$) = 0.55 $\pm$ 0.27(stat). The ratio is close to that  seen by
Kamiokande
and IMB, thus confirming the problem,  but R($\mu/e$) = 1 is also not
excluded: the probability to be in agreement with 1
is about 11\%.

Some remarks are in order.
 The data show the deficit of the $\mu$-like events whereas e-like
events are in a good agreement with predictions. However, for second  0.5
kt*y
only, one finds 22.5 tracks which coincides with predicted value 22.1
track,
i.e., there is no deficit of muon neutrinos. The respective increase of the
number of events could be related to the increase of  mass of the
detector in second series (680 - 750 tons),
and therefore to the increase of the acceptance  to muons.
The double ratio in second series is $R \approx 0.77$  -- more close to 1.
All these features can be a result of statistics and
more data are needed to make a firm conclusion.\\

4. The fluxes of the  upwardgoing muons do not show a deficit of muon
neutrinos.  The corresponding data from
Kamiokande$^{58}$, IMB$^{59}$ and Baksan$^{60}$ are  in  agreement with
predictions.
In particular, the flux measured by
Kamiokande$^{51}$:
$F = (2.04 \pm 0.13) \cdot 10^{-13}$
cm$^{-2}$s$^{-1}$st$^{-1}$
should be compared with expected value
$F = (2.0 - 2.45) \cdot 10^{-13}$
cm$^{-2}$s$^{-1}$st$^{-1}$.
However,
in this analyses the absolute value of the neutrino flux is used which
has an uncertainty up to 30\% . There are new calculations of
$\nu_{\mu}$
-flux at high energies. In$^{61}$ the largest flux has been found
and it is claimed that the deficit of upwardgoing muons  exists too.
On the contrary,  the estimations in$^{62}$ result in  smallest
$\nu_{\mu}$-flux, but in the
same time it is claimed that the uncertainties (following from the $K/\pi$-
ratio, hadron interaction cross sections, neutrino cross-sections) are
underestimated. The realistic value ($\sim$ 40\%) admits different
interpretations of the results.\\

5. The uncertainties in the absolute fluxes are cancelled when one
compares
the number of stopping and through going muons (small uncertainty is
related to energy dependence of flux). The
ratio (stopping/through going) = 0.16 $\pm$ 0.02 measured by IMB$^{59}$
 is in excellent agreement
with prediction: (0.163 $\pm$ 0.05). Since stopping
and through going
muons correspond to different intervals of neutrino energies, the ratio is
sensitive to the energy dependent effects like the neutrino oscillations.\\

6. The deficit of $\nu_{\mu}$-flux can be explained by the
oscillations
$\nu_{\mu} - \nu_e$ or
$\nu_{\mu} - \nu_{\tau}$ or
$\nu_{\mu} - \nu_s$ $^{63}$, where $\nu_{s}$ is the sterile neutrino,
although the last possibility is disfavoured by the nucleosynthesis
consideration.
Negative results give the exclusion region of
the neutrino parameters which however does not cover all the region of
positive results (fig.5).  Such a reconciliation is still possible due to
relatively large error bars in the Frejus data and
large uncertainties in the results on the
upwardgoing muons.
Note that the most conservative limit is
given by Kamiokande, where the angular distribution of the muons was
studied.
The IMB and Baksan strongly restrict the region. Moreover, Baksan data
with  flux calculated by Volkova exclude all the region of the positive
results (but see discussion  in$^{49}$). Also one should mention
the fact that the upwardgoing muons correspond to higher neutrino energies
and  the oscillations $\nu_{\mu} \leftrightarrow \nu_e$ can be suppressed by
matter effect in the
Earth at high energies more strongly than at low energies.
The survival domains are
\beq
\Delta m^2 = (0.3 - 3) \cdot 10^{-2}~ {\rm eV}^2,~~~
\sin^2 2\theta = 0.4 - 0.6
\eeq
for $\nu_{\mu} - \nu_{\tau}$ oscillations  and
\beq
\Delta m^2 = (0.3 - 2) \cdot 10^{-2}~ {\rm eV}^2,~~~
\sin^2 2\theta = 0.35 - 0.8
\eeq
for
$\nu_{\mu} - \nu_e$ (fig.5). In the indicated regions the
data from different experiments are described at  about $2\sigma$
level and, consequently, the total probability that all the
data are fitted by the parameters (17,18) is rather small.

Note that maximal mixing is excluded as a solution of the
atmospheric
neutrino problem by Frejus result on the double ratio and
by IMB result on stopping/through going muons, and the uncertainties of
both results are rather small.
This fact is very important for theoretical implications.\\

7. Another explanation$^{64}$ is related to possible  proton decay:
$p \rightarrow e^+ \nu \nu$. The smallness of the  double
ratio is due to excess of the e-like events from the decay.
The observed value R($\mu/e$) can be reproduced for
the lifetime $T = \left( 4.0^{+1.9}_{-1.0} \right) \cdot 10^{31}$ years.
Obviously, the excess is at energies $E^{vis} < 1$ GeV,  and there
is no deficit of the upwardgoing muons. The energy distribution of events
was in agreement
with distribution observed by Kamiokande. However  the IMB and the  latest
Kamiokande results for $E^{vis} > 1.33$ GeV seem to show the effect
(smallness of double ratio)  at high
energies  too. Also the deficit of the contained events with muon decay
testifies  against proton decay solution.
Moreover, SOUDAN-II can measure a proton recoils due to
the neutrino scattering.
In the analyzed sample 5 showers with a proton recoil have been observed
which is in  agreement with predicted value 3.9.
In case of proton decay there is no proton in final
state.

Larger statistics in the SOUDAN experiment and the calibration of the
water \'{C}erenkov detectors at KEK may change a status of  the problem.\\

\begin{center}
SOLAR NEUTRINOS
\end{center}

Recently, all four collaborations
measuring the
solar neutrino fluxes  have published new results.

1. {\it Homestake experiment}
($\nu_e + ^{37}Cl \rightarrow e + ^{37}Ar$, $E_{th} = 0.816$ MeV).
There are final results from five runs
(115 - 119) and preliminary results from three new runs (120 - 122) of the
measurement of the Ar-production rate$^{65,66}$. The end of the latest run 122
is  dated by 1992.177 y; new points are concentrated around
$N_{Ar} \approx 0.7 - 0.8$ at/day.
The average counting rate over all the time of
observation (after background subtraction) is
\beq
Q_{Ar} = 2.28 \pm 0.16(stat) \pm 0.21(syst)~~ {\rm SNU}.
\eeq
The Ar-production rate in a series of the experiment in 1986 - 1992 (after
stop of the experiment in 1985 - 86), $Q_{Ar} = 2.85 \pm 0.16~$ SNU,
is appreciably higher than the
average. Time combined ($\approx$ 5 years)
Ar-production rate
is shown in fig.6$^{66}$. Clearly the rate in the last bin is higher than
in the previous  ones.
The latest data do not confirm the anticorrelation with solar activity:
large number of the sunspots in 1991 - 1992 was accompanied by high
counting rate.
On the other hand the data confirm 2 - 3 years period variations of
signal.\\

2. {\it Kamiokande III}
($\nu_{e, x} + e \rightarrow \nu_{e,x} + e$, $E_{th} \sim 7.5$ MeV).
The observations during 627
days (Dec '90 - July '93) give
the ratio of the measured flux of the boron
neutrinos to the predicted one
\beq
R_{\nu e}^{III} \equiv \frac{F^{exp}}{F^{SSM}} = 0.54_{-0.05}^{+0.06} \pm
0.06,
\eeq
where $F^{SSM} \equiv 5.8 \cdot 10^5$ cm$^{-1}$ s$^{-1}$ is the central
value of the
flux predicted by the Standard Solar Model (SSM) of Bahcall and
Pinsonneault$^{73}$.  The combined result from Kamiokande-II and III is
\beq
R_{\nu e}^{II + III} = 0.50 \pm 0.04 (stat.) \pm 0.06 (syst.), ~~~ (1
\sigma).
\eeq
Time dependence of signals is shown in fig.7. The data  agree
with constant neutrino flux. No anticorrelations with solar activity
was found. Possible time variations should not exceed  30\%.
The energy distribution of the events can be fitted with practically the
same  probabilities by constant and MSW-nonadiabatic suppression factors.\\

3. {\it GALLEX}$^{68,69}$
($\nu_e + ^{71}Ga \rightarrow e + ^{71}Ge$, $E_{th} = 0.233$ MeV).
Final results from 15 runs of GALLEX-I experiment
have been  published$^{69}$.
Exposure period is from 14 May 1991 to 29 April 1992; counting has been
finished on November 1992. The average Ge-production rate
is just 2 SNU below the preliminary result, see (22)
(although the changes of the results of the individual runs are larger).

There is a number of changes in the GALLEX-II experiment: it runs in another
tank, the exposure time is larger: about 1 month, etc. .
Preliminary results from 6  runs of the GALLEX-II (19 August 1992 - 3
February 1993) give the average  Ge-production rate, $Q_{Ge}^{II}$,
 about 1$\sigma$ higher than in first series.
The points have rather small spread around 100 SNU, and only in one run
a low signal has been detected.
Combined result of GALLEX-I and GALLEX-II is 87 SNU.
\beq
\begin{tabular}{|l|l|}
\hline
\sl GALLEX   & \sl $Q_{Ge}$,~ {\rm SNU}~~ ($1\sigma$) \\
\hline
$Q_{Ge}^I$ & 81 $\pm$ 17(stat) $\pm$ 9(syst)  \\
$Q_{Ge}^{II}$ &97 $\pm$ 23(stat) $\pm$ 7(syst)~~ {\rm (prelim.)} \\
$Q_{Ge}^{I+II}$ & 87 $\pm$ 14(stat) $\pm$ 7(syst)~~({\rm prelim.)} \\
\hline
\end{tabular}
\eeq
The following  aspects of new result are important for implications.
1). The error bars become smaller: a combined error is $\sim$ 16 SNU
as compared with 24 SNU in the first series.
2). The lower limit goes up:
\beq
Q_{Ge} >  \left\{ \matrix{ 71~ {\rm SNU} ~~ 1 \sigma \cr
                           56~ {\rm SNU} ~~ 2 \sigma
\cr  } \right. .
\eeq
3). The deviation from the SSM predictions (122 - 132 SNU) is on the same
level as before:
the data are $\sim (2.2 - 2.6)\sigma$ below the expected value.

Let us note that the
production rate obtained from
 L-peak (1.2 kev) is larger than  that from K-peak (10.4 kev) in both
series:
\beq
\begin{tabular}{|l|l|l|}
\hline
\sl GALLEX & \sl  K-peak  & \sl L-peak  \\
\hline
I      &   64 ~SNU            & 105~SNU \\
II     &   89 ~SNU            & 110~SNU \\
\hline
\end{tabular}.
\eeq
(Obviously, the signals from both peaks should be the same). The
above result may be just a statistical fluctuation (at present of the order
of 2$\sigma$) or indication on some systematical error. (The
background at low energies is larger).

At present GALLEX-II experiment is performed with a frequency 1 run
a month. The calibration  with $^{51}Cr$ source is planning to be in
summer 1994.\\

4. {\it SAGE}.  The preliminary result$^{70}$ from runs of 1990 and 1991 was
$Q_{Ge} = 58 \pm 14(stat) \pm 7(syst)~ SNU ~~(1\sigma)$.
The results of 4 new runs have been published$^{71}$ at TAUP-93.
In two runs best fit corresponds to  zero signal.  In  combined statistical
analyses the
run 1990-5 is again removed and
combined Ge-production rate in 15 runs is
\beq
Q_{Ge} = 70 \pm 19(stat) \pm 10(syst)~ SNU ~~(prelim.~~1\sigma).
\eeq
Let us remark the following.
In 5 runs (from 15)  best fit gives zero flux. The goodness of the
fit is
apparently lower in the runs with low signal. Time distribution of
events in the counters (summed over all the runs) agrees with 16.5 day
meanlife, but
it does not yet give a compelling proof that signal corresponds to the
$^{71}$Ge-decay. It seems that the distribution can be fitted  by a decay
curve with
shorter period. Although the systematic error due to radon is estimated to be
rather small (5 SNU).\\

5. There are {\it two aspect} of the solar neutrino problem.

1). All the experiments have  detected signals  which are lower
than the predictions of the consistent solar models (fig. 8); the ratios
R $\equiv$ (observations)/(central predicted values)  for
two SSM  are
\beq
\begin{tabular}{|l|l|l|}
\hline
\sl R     & \sl  $B-P^{73}$        & \sl  $TC - L^{74}$  \\
\hline
$R_{Ar}$    & 0.285 $\pm$ 0.030      &  0.365 $\pm$ 0.030 \\
$R_{\nu e}$ & 0.50  $\pm$ 0.07       &  0.63  $\pm$ 0.07  \\
$R_{Ge}$    & 0.66  $\pm$ 0.12       &  0.71  $\pm$ 0.13 \\
\hline
\end{tabular}
\eeq
(The errors are only from the experiment).

2).  Signal in the Cl - Ar experiment is suppressed more strongly than the
signal in the Kamiokande. It follows from (26) that
\beq
\frac{R_{Ar}}{R_{\nu e}} = \left\{ \matrix{ 0.58 \pm 0.12, ~~ B-P \cr
                                       0.57 \pm 0.12, ~~ TC-L   \cr
} \right .  .
\eeq
This statement can be relaxed if one takes into account the Cl- Ar data only
for a period of the operation of Kamiokande:
$\frac{R_{Ar}}{R_{\nu e}} = 0.78 \pm 0.22~$. But this evidently, implies
the time variations of Homestake data.

One can  perform a direct test of  consistency of
Cl - Ar and Kamiokande results suggesting that there is no distortion of the
energy spectrum and that  Kamiokande signal is due to the electron
neutrino  scattering only.
Taking the boron neutrino flux as measured
by Kamiokande one finds the contribution of boron
neutrinos to  Ar production rate:
$Q_{Ar}^{B} = 3.0 \pm 0.4$ SNU which is even larger than total  measured
rate even if one neglects the contribution from Be-neutrinos.
With Cl-data during 1986 - 1992 one removes a direct contradiction,
but it is difficult to reproduce such a situation by  modification of
the solar model.

Tacking into account a difference in the
thresholds of different experiments one can conclude on the following
energy dependence of the suppression factors:
there is weak (or zero) suppression at low energies
(pp-neutrinos), strong suppression at moderate energies (Be- neutrinos);
moderate suppression at high energies (B-neutrinos). \\

There are several directions in which the source of the discrepancy is
looked for.

6. After GALLEX publications it becomes fashionable to discuss the
``detection" solution of the problem, keeping in mind that some of the
experimental results may have  incorrect interpretation.

GALLEX results are rather stable and
convincing. It is difficult to
expect  appreciable changes of numbers. Probably, fixing of the
K - L difference diminishes the counting rate. On the other hand
the calibration experiment may result in the renormalization of the effect
and in  increase of the measured neutrino flux. SAGE experiment confirms
GALLEX results.

Kamiokande results are stable, convincing, and the experiment had been
calibrated.

Homestake experiment
shows the strongest suppression of signal.
There is no calibration.
The following features of the data which have small statistical
probability are remarked:
the probability that the data correspond to the constant flux is smaller
than 5\%$^{65}$. The suggested effect of the anticorrelation with solar
activity
is now $\sim 2 \sigma$. There is very low signal during 1978 (five runs
with near to zero counting rate). There is a general tendency in
increase of the signal. The signal during 1986 - 1992 is appreciably larger
than the average one.
There is a concentration of the points around $N_{Ar} \sim 0.7 - 0.8$
 at/day. The signal at the level $Q_{Ar} \sim 4$ SNU could be accommodated
by astrophysics, thus changing the status of the problem.\\

7. {\it Astrophysical solution}.
Neutrino fluxes decrease with diminishing of temperatures in the center of
the Sun.
A number of  modifications of  solar models
(such as low concentration of heavy elements, low opacity,
WIMPS, fast central rotation etc.) were suggested which result in
$T_c$ decrease. However,  diminishing $T_c$ suppresses the boron neutrino
flux  more strongly than the berillium neutrino flux,
(using 1000SSM the empirical relations  $F_{B} \propto T^{18}_c$ and
$F_{Be} \propto T^{8}_c$ have been found$^{72}$ for small region
$T_c$ near $T^{SSM}_c$)
and consequently, the double ratio in (27) should be larger than 1 in
contradiction with experimental result. (Although, the production regions
of boron and berillium neutrinos differs and modifying the temperature
profile one  may change, to some extend, the above relation$^{76,77}$
). Essentially
for this
reason a combined fit of all the data for arbitrary astrophysical
parameters is rather bad (fig. 9) (any astrophysical solutions
are  excluded at 98\% C.L.$^{78}$).
Moreover, this bad ``best
fit"
can be  reached at unacceptably strong stretching of the model (e.g., by
diminishing central temperatures by 6 - 8 \%)$^{78}$.

This statement is relaxed if one uses the Homestake data only
for a period of the operation of Kamiokande. But this implies the
time variations of the Cl - Ar signal.\\

8. {\it Particle physics solution}.
The data obtained so far can be perfectly described by the
resonant flavor conversion (MSW-effect)$^{79}$ $\nu_e \rightarrow
\nu_{\mu} (\nu_{\tau})$ . Energy dependence of the suppression factor
allows to reproduce at definite values of parameters even the central
values
of the observed signals. For two neutrino mixing (which is a good
approximation
when the mass spectrum has a strong hierarchy and the admixture of the third
neutrino is small) the data pick up two regions of parameters (see fig.
10). The one corresponds to  small (vacuum) mixing
solutions$^{80,81}$:
\beq
\Delta m^2 = (0.5 -1.2) \cdot 10^{-5}~ {\rm eV}^2,~~~
\sin^2 2\theta = (0.3 - 1.0) \cdot 10^{-2} ,
\eeq
another one to large mixing solution:
\beq
\Delta m^2 = (1 - 3) \cdot 10^{-5}~ {\rm eV}^2,~~~
\sin^2 2\theta = (0.65 - 0.85).
\eeq
Lower bounds on Ge-production rate from  GALLEX (24)
disfavor the
third region with large mixing and small $\Delta m^2$.
In presense of  third
neutrino the  allowed domains become larger, in particular,  the
region of  small mixing solutions can be extended$^{82}$ up to
$\sin^2 2\theta = 8 \cdot 10^{-4}$ and
$\Delta m^2 = 8 \cdot 10^{-5}~ {\rm eV}^2$. \\

9. The region of small mixing is quite plausible from  theoretical point
of view.
The angles in (28) are
a little bit smaller than
$
\theta_l \equiv  \sqrt {\frac{m_e}{m_{\mu}}} ,
$
where $m_e$ and  $m_{\mu}$ are the  masses of the electron  and muon
($\sin^2 2\theta_l \sim 0.02$).
One can correct this expression by adding the contribution from the
neutrino mass ratio ($\theta_{\nu}$):
\beq
\theta_{e \mu} = \left |\sqrt \frac{m_e}{m_{\mu}} - e^{i\phi}
\theta_{\nu} \right |,
\eeq
where $\phi$ is a phase. Such a relation between the angles and the masses
is similar to the relation in  quark sector$^{83, 84}$ and
follows naturally from Fritzsch ansatz for mass matrices.
There are the models which realize such a possibility in terms of the
see-saw mechanism of mass generation$^{85}$.

Also  large lepton mixing corresponding to solutions (29) can
be reproduced by the see-saw mechanism. The enhancement of mixing (as
compared with quark sector) may take place due to certain structure of
mass matrix of the right handed neutrinos which, in turn, can be explained
by certain family symmetry$^{86}$. \\

10. Alternatively the data can be described by {\it long length vacuum
oscillations}  (``just-so")$^{87}$ with parameters$^{88, 89}$:
\beq
\Delta m^2 = (0.5 - 1.0) \cdot 10^{-10}~ {\rm eV}^2,
{}~~~
\sin^2 2\theta = 0.70 - 1.0,
\eeq
(see fig. 11). The parameters (31) can be rather naturally
reproduced by Planck scale interactions with  flavor universal
effective couplings$^{90,91}$.
However, the mixing with parameters (31) is disfavoured by the data
from SN1987A$^{92}$.
For large mixing the transitions $\bar{\nu}_{\mu}
\leftrightarrow \bar{\nu}_e$
($\bar{\nu}_{\tau}$) result in the modification of the
$\bar{\nu}_e$-energy spectrum.
In particular, the  appearance of the high energy tail is expected,
since  the original $\bar{\nu}_{\mu}$
($\bar{\nu}_{\tau}$)
energy spectrum
has a larger average energy than the spectrum of $\bar{\nu}_e$.
The events with E $>$ 40 -  50 MeV are predicted in contrast
with observations.
The excluded region (fig. 11) covers the region of
``just-so" solution$^{92}$. \\

Although a solution of the solar neutrino problem will be possible with
data from new solar neutrino experiments, already present data allow to
make some firm conclusions (independent on the model of the Sun etc.).
Kamiokande gives the model independent restrictions on the neutrino
parameters from measurements of energy spectra of the events and from search
for day/night effect$^{67}$.
Lower bound on $Q_{Ge}$  obtained by GALLEX  allows to
exclude (practically in model independent way) a large region of the neutrino
parameters (fig. 10). \\

12. {\it Reconciliation}. Is it possible to reconcile the particle
physics solution of the solar neutrino problem (28,29,31) with other
positive indications of nonzero neutrino masses and mixing, namely,
with explanation  of muon neutrino deficit in the atmospheric flux
in terms of oscillations or/and
with existence of neutrino with  mass in the cosmologically
interesting region
\beq
m \sim 2 - 7~~ {\rm eV} .
\eeq
Such a neutrino could be a component of the hot dark matter which is
needed to explain
a formation of  large scale structure of the Universe$^{93}$ .

Let us make remarks.

1). The neutrino parameters (28)
for $\nu_e \rightarrow \nu_{\mu}$ can be easily
reproduced in the see-saw mechanism with Majorana mass of the RH
neutrinos $10^{10} - 10^{12}$ GeV. For third neutrino, being main
component of $\nu_{\tau}$, one gets $m_3 \sim 1 - 30$ eV just in the region
(32). New experiments at CERN (CHORUS and NOMAD) will be able to study a
large region of mixing angles of such a neutrino. However, in this case
there is no room for a solution of the atmospheric neutrino problem.

2). The parameters for the solar and atmospheric neutrino problems can
be reconciled in terms of the see-saw mechanism if $m_2 \sim
m_{\odot}$ and $m_3 \sim 0.1$ eV. Moreover, large mixing of
$\nu_e$ and $\nu_{\mu}$ may be related to a relatively weak hierarchy of
masses: $m_3/m_2 = 10 - 30$. However, in this scenario  neutrinos can not
play the role  of the hot dark matter (32).

3). The reconciliation of the solutions of all three problems requires
more sophisticated models. One possibility is the highly degenerate
spectrum with $m_1 \approx m_2 \approx m_3 \approx 1 - 2$ eV $^{94}$.
Small mass splitting: $(m_2 - m_1)/m_2 \sim 10^{-5}$ and $(m_3 - m_2)/m_2
\sim 10^{-3}$ can solve the solar and the atmospheric problems
correspondently.
Another possibility is related to the introduction of the light
sterile neutrino(s) which could play the role of the hot dark matter or
participate in the conversion of solar neutrinos$^{95}$.\\

\begin{center}
 CONCLUSIONS
\end{center}

The laboratory experiments (direct measurements of the neutrino masses,
searches for the double beta decay, kinks on the Kurie plots, oscillations,
neutrino decays, etc.) give  negative results: no effects of the
neutrino masses, mixing as well as new interactions have been found. This
gives  the  upper bounds on the masses and mixing
at the level of  predictions of the ``electroweak see-saw".

Positive indications of the existence of the neutrino masses and mixing
are related to the atmospheric and  solar neutrino problems. Recent data do
not change essentially their status. New Kamiokande data
further confirm the smallness of the double ratio.
First ``provisional" results from SOUDAN-II  are not yet decisive,
although they
indicate the smallness of the double ratio too. The neutrino oscillations
are considered as the most plausible solution of the problem.

New solar neutrino data also confirm previous results. If the
interpretation of all the experimental data is correct  any
astrophysical solutions of the problem are disfavou-\\
red. The results can be
well described in terms of the resonant conversion or long length vacuum
oscillations (although the latter is disfavoured by SN1987A data).

Reconciliation of these results allows to trace  possible patterns of the
neutrino masses and lepton mixing, and the latter  seems not to coincide
with  mixing in quark sector.\\

\begin{center}
ACKNOWLEDGEMENTS
\end{center}

I would like to thank A. Barabash, J. Bonn, R. Frosch,
K. Lande, F. Laplanche,  P. Liechfield, T. Nakada, J. Panman, A. Piepke, S. P.
Mikheyev,   G.~Senjanovi\'{c},
for valuable discussions and for the help in preparation of
this talk.

The participation in the Symposium was supported by the Soros
Foundation and the American Physical Society.\\

\begin{center}
 REFERENCES
\end{center}

\noindent
1. M. Swartz, these Proceedings, G. Coignet, these Proceedings.\\
\noindent
2. Los Alamos: R.~G.~H.~Robertson et al., Phys. Rev. Lett. {\bf 67}
(1991) 957. \\
\noindent
3. Mainz: Ch.~Weinheimer et al., Phys. Lett. {\bf B300} (1993) 210.\\
\noindent
4. Livermore: W.~Stoeffl, Bull. Am. Phys. Soc. {\bf 37} (1992) 925.\\
\noindent
5. Zurich: Holtschuh et al., Phys.~Lett. {\bf B287} (1992) 381. \\
\noindent
6. J. Bonn, private communication.\\
\noindent
7. Tokyo: H. Kawakami et al., Phys. Lett. {\bf B256} (1991) 105.\\
\noindent
8. J. Wilkerson, Nucl. Phys. B, (Proc. and Suppl.) {\bf 31} 32.\\
\noindent
9. M.~Daum et al., Phys. Lett. {\bf B 265} (1991) 425.\\
\noindent
10. M. Janousch et al., Proc. of the
Low Energy Muon Science meeting, April 1993,\\
\phantom{ujh}Santa Fe, New Mexico, USA.\\
\noindent
11. ARGUS: H.~Albrecht et al., Phys. Lett. {\bf B202} (1992) 224.\\
\noindent
12. CLEO: D.~Cinabro et al., Phys. Rev. Lett. {\bf 70} (1993) 3700.\\
\noindent
13. E.~W.~Kolb, M.~S.~Turner, A.~Chakavorty and D.~N.~Schramm, Phys. Rev.
Lett.\\
\phantom{tghy}{\bf 67} (1991) 533.\\
\noindent
14. A.~D.~Dolgov and I.~Z.~Rothstein, Phys.~Rev.~Lett. {\bf 71} (1993) 476.\\
\noindent
15. M.~Kawasaki, G.~Steigman, H.-S.~Kang,
Nucl. Phys. {\bf B403} (1993) 671. \\
\noindent
16. T. Walker, private communication,  G. Steigman et al., in
preparation.\\
\noindent
17. Heidelberg-Moscow: A.~Balysh  et al., Phys. Lett. {\bf B283} (1992) 32;
A.~Piepke, talk\\
\phantom{hyt} given at
the International Europhysics Conference on High
Energy Physics, Mar-\\
\phantom{jgres}seille, July  22 - 28 (1993). \\
\noindent
18. M. Beck et al., Phys. Rev. Lett., {\bf 70} (1993) 2853. \\
\noindent
19. A. Piepke, private communication.\\
\noindent
20. J.-C.Vuilleumier et al., Phys. Rev., {\bf D48} (1993) 1009.\\
\noindent
21. NEMO Collaboration: R. Arnold et al., Proc. of
the International Europhysics\\
\phantom{res} Conference on High Energy Physics, Marseille,
July
22 - 28 (1993). \\
\noindent
23. M. Alston-Garnjost  et al., Phys. Rev. Lett., {\bf 71} (1993) 831.\\
\noindent
24. T.~Bernatowicz et al., Phys.~Rev.~Lett., {\bf 69} (1992) 2341. \\
\noindent
25. IGEX Collaboration: A.~Morales in: TAUP'91, eds. A.Bottino, P
Monacelli, p. 97,\\
\phantom{hgt}Editions Frontieres 1989.\\
\noindent
26. S. T. Petcov and A. Yu. Smirnov, Preprint IC/93/360 and
SISSA 113/93/EP.\\
\noindent
27. S. Chong and K. Choi, Preprint SNUTP 92-87 (1993).\\
\noindent
28. S. R. Elliott et al., Nucl. Phys. B, (Proc. Suppl.) {\bf 31} (1993) 68.\\
\noindent
29. Z. G. Berezhiani, A. Yu. Smirnov, and J. W. F. Valle, Phys. Lett.
{\bf 291B} (1992)\\
\phantom{ytgf} 99.\\
\noindent
30. Z. G. Berezhiani (unpublished);\\
\phantom{yhat}C. P. Burgess and O. F. Hern\'andez, Phys. Rev. {\bf 48} (1993)
4326.\\
\noindent
31. C. P. Burgess and J. M. Cline, Phys. Lett. {\bf 298B} (1993) 141;
Preprint McGill/93-\\
\phantom{ytgf}02, TPI-MINN-93/28-T, UMN-TH-1204-93.\\
\noindent
32. D. Grasso, M. Lusignoli  and M. Roncadelli, Phys. Lett. {\bf B288}
(1992) 140. \\
\noindent
33. E. Kh. Akhmedov, Z. G. Berezhiani, R. N. Mohapatra and G. Senjanovi\'c,
Phys.\\
\phantom{ytgf}Lett. {\bf B299} (1993) 90.  \\
\noindent
34. T.~Ohshima et al., Phys.~Rev. {\bf D47} (1993) 4840.\\
\noindent
35. E.~Holzschuh and W.~Kundig, in Perspectives in Neutrinos, Atomic
Physics, and \\
\phantom{uytr}Gravitation, XIII Moriond Workshop, Villars-sur-Ollon,
Switzerland, January\\
\phantom{tyui}30 - February 6 (1993).\\
\noindent
36. J.~L.~Mortara et al., Phys. Rev. Lett. {\bf 70} (1992) 394.\\
\noindent
37. G.~E.~Berman et al., to submitted to Phys. Rev. Lett. .\\
\noindent
38. M.~Bahran and G.~R.~Kalbfleisch, Phys.~Lett. {\bf B303} (1993) 355;
Phys.~Rev. {\bf D47} \\
\phantom{opi}(1993) R754.\\
\noindent
39. A. Hime,
5th International Workshop on Neutrino Telescopes, ed. by
M. Baldo \\
\phantom{ytgf}Ceolin, (1993) p.359.\\
\noindent
40. L. Oberauer, talk given at
the Int. Europhysics Conference on High\\
\phantom{res} July
22 - 28 (1993). \\
\noindent
41. A. Griffols and E. Masso, Phys. Lett., {\bf B 242} (1990) 67.\\
\phantom{fgtr}R. Gandhi and A. Burrows, Phys. Lett. {\bf B246} (1990) 149.\\
\noindent
42. S. Shevchenko, talk given at
the Int. Europhysics Conference on High Energy\\
\phantom{res} Physics, Marseille, July  22 - 28 (1993). \\
\noindent
43. H.~Pessard, talk given at the International Conference
Neutral Currents: Twenty
\\
\phantom{rfgt}
Years Later, Paris, France, July 6 - 9, 1993.\\
\noindent
44. CHARM: M. Gruw\'e et al., CERN-PPE/93-93. \\
\noindent
45. E645: S. J. Freedman et al., Phys. Rev.{\bf D47} (1993) 811. \\
\noindent
46. CHORUS: N. Armenise et al., CERN-SPSC/90-42 (1990).\\
\noindent
47. NOMAD: P. Astier et al., CERN-SPSLC/92-51 (1992).\\
\noindent
48. Y.-Z. Qian et al., Phys. Rev. Lett. {\bf 71} (1993) 1965.\\
\noindent
49. Y.~Totsuka, Nucl.~Phys.~B (Proc. Suppl.) {\bf 31} (1993) 428.\\
\noindent
50. Kamiokande: K.S. Hirata et al., Phys. Lett. {\bf 280B} (1992) 146;
T. Kajita, Proc. \\
\phantom{rfgt}
of the International symposium "Fronties of Neutrino
Astrophysics" Ed. by \\
\phantom{rfgt}
Y. Suzuki and K. Nakamura, Universal Academy Press,
Inc. - Tokyo, Japan,\\
\phantom{rfgt}
 p. 293. \\
\noindent
51. Y. Suzuki, talk given at the Workshop TAUP-93.\\
\noindent
52. IMB: D. Casper et al., Phys. Rev. Lett. {\bf 66} (1991) 2561.\\
\noindent
53. K. Nishikawa et al., KEK Preprint 93-55, INS-Rep.-991,
ICRR-Report-297-93-9. \\
\noindent
54. Frejus: Ch. Berger et al., Phys. Lett. {\bf B227} 489; ibidem
{\bf B245} (1990) 305.\\
\noindent
55. NUSEX: M. A. Aglietta et al., Europhys. Lett. {\bf 8} (1989) 611.\\
\noindent
56. SOUDAN-2: E. Peterson, contributed paper \# 98.\\
\noindent
57. P. J. Litchfield, Proc. of the  5th
Int. Workshop on Neutrino
Telescopes, ed. by\\
\phantom{fgty}M. Baldo-Ceolin, p. 235 (1993); and in
Int. Europhysics Conference on High\\
\phantom{oikl} Energy Physics,  Marseille, July  22 - 28 (1993).\\
\noindent
58. Kamiokande: Y. Oyama et al., Phys. Rev. {\bf D39} (1989) 1481;
Y. Suzuki, see ref.$^{51}$\\
\noindent
59. IMB: R. Becker-Szendy et al., Phys. Rev. Lett. {\bf 69} (1992) 1010.\\
\noindent
60. Baksan: M. M. Boliev et al., Proc. of Third Int.
International Workshop on Neu\\
\phantom{rfgt} trino Telescopes, ed. by  M. Baldo Ceolin, p.235 (1991). \\
\noindent
61. W. Frati, T. K. Gaisser, A. K. Mann and T. Stanev, Phys. Rev.,
{\bf D48} (1993) 1140.\\ \noindent
62. D. Perkins, Nucl. Phys., {\bf B399} (1993) 3.\\
\noindent
63. E. Kh. Akhmedov, P. Lipari and M. Lusignoli, Phys. Lett. {\bf B300}
(1993) 128.\\
\noindent
64. W. A. Mann, T. Kafka and W. Leeson, Phys. Lett. {\bf B291} (1992)
200.\\
\noindent
65. R. Davis Jr., Proc of the 23th ICRC, Calgary, Canada (1993).\\
\noindent
66. K. Lande, private communication.\\
\noindent
67. Kamiokande: K. S. Hirata et al Phys. Rev. Lett. {\bf 66} (1991) 9;
Phys. Rev. {\bf D44}\\
\phantom{rfgt} (1991) 2241; Y. Suzuki see ref.$^{38}$.\\
\noindent
68. GALLEX-I: P. Anselmann et al., Phys. Lett. {\bf B285} (1992) 376, 390.\\
\noindent
69. P. Anselmann et al., Phys. Lett. {\bf B314} (1993) 445.\\
\noindent
70. SAGE: A. I. Abasov et al., Phys. Rev. Lett. {\bf 67} (1991) 3332;
G. Zatsepin,  Proc.\\
\phantom{mnbh} of the Int Symposium "Frontiers of Neutrino
Astrophysics", Ed. by Y. Suzuki\\
\phantom{ijnb}and K. Nakamura, Universal Academy Press,
Inc. (1993) p. 71.\\
\noindent
71. V. N. Gavrin, Talk given at Int. Workshop TAUP-93.\\
\noindent
72. J. N. Bahcall and R. K. Ulrich Rev. Mod. Phys. {\bf 60} (1989)
297.\\ \noindent
73. J. N. Bahcall and M. M. Pinsonneault, Rev. Mod. Phys. {\bf 64} (1992)
885.\\
\noindent
74. S. Turck-Chieze and I. Lopez, Astroph. J., {\bf 408} (1993) 347.\\
\noindent
75. G. Bertomieu, J. Provost and J. Morel, Astron. Astroph. {\bf 268} (1993)
775.\\ \noindent
76. V.Castellani, S.Degl'Innocenti, G. Fiorentini, Phys. Lett. {\bf 303B}
(1993) 68.\\
\noindent
77. A. Bertin et al., Phys. Lett. {\bf B303} (1993) 81.\\
\noindent
78. S. Bludman et al., Preprint UPR-0572T (contributed paper \# 114).\\
\noindent
79. S.~P.~Mikheyev and A.~Yu.Smirnov, Sov.~J.~Nucl.~
Phys.~{\bf 42 }(1985) 913; Sov.\\
\phantom{agtd}Phys.~JETP {\bf 64} (1986) 4;
 L.~Wolfenstein, Phys.~Rev.~
 D {\bf 17}, 2369 (1978), ibidem,\\
\phantom{yuij}{\bf D20} 2634 (1979).\\
\noindent
80. P.~I.~Krastev and
S.~T.~Petcov, Phys.~Lett. {\bf B299},
 (1993) 99;
 L.~M.~Krauss,\\
\phantom{yujh}E.~Gates and
M.~White, Phys.~Lett. {\bf B298},
 (1993) 94,
Phys.~Rev.~Lett., {\bf 70},\\
\phantom{olki} (1993) 375,
S.~A.~Bludman et al., Phys. Rev. {\bf D47} (1993) 2220; N.~Hata and\\
\phantom{yhgt}P.~Langacker, Univ. of Pennsylvania preprint UPR-0570T;
G.~L.~Fogli,  E.~Lisi\\
\phantom{olki}and D.~Montanino, Preprint CERN-TH 6944/93,
BARI-TH/146-93. \\
\noindent
81. N. Hata and P. Langacker, contributed paper \# 307.\\
\noindent
82. X.~Shi,  D.~N.~Schramm, and
J.~N.~Bahcall, Phys.~Rev.~Lett.~{\bf 69} (1992) 717;\\
\phantom{ert} A.~Yu.~Smirnov, ICTP preprint IC/92/429;
D.~Harley, T.~K.~Kuo and J.~Pan-\\
\phantom{olki}taleone, Phys.~Rev. {\bf D47} (1993)
4059;
A.~S.~Joshipura and P.~I.~Krastev,\\
\phantom{tgvh}Preprint IFP-472-UNC, PRL-TH-93/13.\\
\noindent
83. S.~Weinberg, ``A  Festschrift for I.~I.~Raby",
Transaction of the New York Academy\\
\phantom{ikju} of Sci. {\bf 38} (1977) 185; F.~Wilczek and
A.~Zee, Phys.~Lett. {\bf B70} (1977)  418.\\
\noindent
84. H.~Fritzsch, Phys.~Lett. {\bf B70} (1977) 436.\\
\noindent
85. A. Yu. Smirnov, talk given at Int. Conf.:
``Neutral Currents: twenty years later",\\
\phantom{lmnb} Preprint IC/93/259, and references therein.\\
\noindent
86. A.~Yu.~Smirnov, Phys. Rev. {\bf D48} (1993) 3264.\\
\noindent
87. V.~N.~Gribov and B.~M.~Pontecorvo, Phys.~Lett. {\bf 28}, (1967) 493;
J.~N.~Bahcall \\
\phantom{poil}and  S.~C.~Frautschi, Phys.~Lett. {\bf B29}, (1969) 623;
V.~Barger, R.~J.~N.~Phillips,\\
\phantom{ploi}and K.~Whisnant, Phys.~Rev. {\bf D24},
(1981) 538;
S.~L.~Glashow and L.~M.~Krauss,\\
\phantom{oiuk}Phys.~Lett. {\bf B190},  (1987) 199.\\
\noindent
88.  see for recent analysis  P.~I.~Krastev and S.~T.~Petcov, Phys.~Lett.
{\bf  B285},\\
\phantom{yuih}(1992) 85.\\
\noindent
89. S. T. Petcov, talk given at
the Int. Europhysics Conference on High Energy\\
\phantom{mres}Physics, Marseille, July 22 - 28 (1993). \\
\noindent
90. R.~Barbieri, J.~Ellis and M.~K.~Gaillard, Phys. Lett. {\bf B90} (1980)
249.\\
\noindent
91. E.~Kh.~Akhmedov, Z.~G.~Berezhiani and G.~Senjanovi\`{c}, Phys.~Rev.~Lett.,
{\bf 69}\\
\phantom{ikln} (1992) 3013.\\
\noindent
92. A.~Yu.~Smirnov, D.~N.~Spergel and J.~N.~Bahcall,
Preprint IASSNS-AST 93/15 \\
\phantom{oiuj}(to be published in Phys. Rev. D). \\
\noindent
93. E.~L.~Wright et al., Astroph.J. {\bf 396} (1992) L13; P.~Davis,
F.~J.~Summers
and\\
\phantom{vfgb} D.~Schlegel, Nature {\bf 359} (1992) 393;
R.~K.~Schafer and Q.~Shafi, Bartol preprint\\
\phantom{iuyt} BA-92-28, (1992);
J.~A.~Holzman and J.~R.~Primack, Astroph.J. {\bf 405}\\
\phantom{plkj} (1993) 428.\\
\noindent
94. D. O. Caldwell and R. N. Mohapatra, Phys. Rev. {\bf D48} (1993) 3259.\\
\noindent
95. J.T. Peltoniemi and J.W.F. Valle, contributed papers \# 210, 211,
Nucl.Phys. {\bf B406}\\
\phantom{oiuy}(1993) 409.\\

\newpage

\begin{center}
FIGURE CAPTIONS
\end{center}

Fig. 1.  The bounds on the electron (anti)
neutrino mass from tritium $\beta$-decay
experiments: Los Alamos$^{2}$, Mainz$^{3}$,
Livermore$^{4}$, Z\"urich$^{5}$, Tokyo$^{7}$, (95\% CL).
Also the upper bound on the effective
Majorana mass of the electron neutrino from
double beta decay  searches is shown.\\

Fig. 2. The restrictions on the tau neutrino Majorana mass as functions
of the lifetime (invisible decay $\nu_{\tau} \rightarrow \nu' \chi$ is
suggested). Hatched line shows the laboratory bounds from ARGUS and
CLEO. Solid lines correspond to restrictions from the Primordial
nucleosynthesis:
1 - from ref.$^{13}$,
2 - from ref.$^{14}$
3 - from ref.$^{16}$. Dashed line shows the bound from total energy
density in the Universe.\\

Fig. 3.  Results from direct searches for the neutrino mixing.
Hatched lines show present limits from the oscillation experiments
($\nu_e \leftarrow \nu_x$,
$\nu_e \leftrightarrow \nu_{\tau}$). Dashed line shows the provisional
result from the BUGEY-III experiment; dotted lines correspond to the level
of sensitivity of future experiments ($\nu_e \leftrightarrow \nu_{\tau}$);
Dashed dotted lines show the upper bound from double beta decay searches
($m_{ee}$ = 1.4 eV)
as well as the level of the sensitivity of future searches
($m_{ee}$ = 0.1 eV). Also shown are the bounds on mixing from searches for
kinks in Kurie plots in the kev - region:
1 - INS, 2 - Z\"urich, 3 - Argonne,
4 - Oklahoma.\\

Fig. 4.  Results on double ratio for atmospheric neutrinos (1$\sigma$).\\

Fig. 5.  Allowed regions of oscillation parameters for atmospheric neutrinos
(shadowed)
The restrictions follow from oscillations
experiments at reactors and accelerators (hatched lines),  contained
events (solid lines), upward going muons (dashed lines), stopping upward
going muons (dashed-dotted line). (See$^{49,63}$ for details).\\

Fig. 6.  Time averaged signal in the Cl - Ar experiment on the solar
neutrinos$^{66}$. The expected counting rate according to SSM is about 1.6
at/day. \\

Fig. 7.  Results from Kamiokande II and III on solar neutrinos
(from$^{51}$).\\

Fig. 8.  Comparison of the observed signals
(hatched regions) with predictions of different standard
solar models ($1\sigma$): 1, 2 - Bahcall-Pinsonneault$^{73}$ (with and
 without diffusion), 3 - Turck-Chieze-Lopez$^{74}$
4 - Bertomieu et al.$^{75}$, 5 - Castellani et al.$^{76}$.\\

Fig. 9.  The allowed regions of the boron and berillium neutrino fluxes from
the combined fit of the Kamiokande, Homestake and Gallium results at
90, 95 and 99\% C.L. (negative values of $\phi(Be)$ are allowed).
For $\phi(Be) > 0$, $\chi^2 > 5.6 $, i.e. any astrophysical solution are
excluded at $> 98\%$ C.L. . Also shown are the predictions from various
nonstandard solar models (from paper $^{78}$).\\

Fig. 10. The MSW solutions of the solar neutrino problem. The allowed
regions of neutrino parameters in two neutrino case (from$^{81}$).
Also the region excluded by lower limit from GALLEX experiment is shown.\\

Fig. 11. Vacuum oscillation (``just-so") solution of the solar
neutrino problem. The regions of the parameters for two
neutrino mixing $\nu_e \leftrightarrow \nu_{\mu}$
(from$^{88,89}$). Also shown are the upper bound on the
mixing from SN1987A (dashed line)$^{92}$ and the predictions from flavor
universal Planck scale interaction$^{91}$.

\end{document}